\documentclass[%
 reprint,
superscriptaddress,
 amsmath,amssymb,
 aps,
 prl,
]{revtex4-2}

\usepackage{graphicx}% Include figure files
\usepackage{dcolumn}% Align table columns on decimal point
\usepackage{bm}% bold math
\usepackage{color}
\begin{document}

\preprint{APS/123-QED}

\title{Photonic Stochastic Emergent Storage: Exploiting Scattering-intrinsic Patterns for Programmable Deep Classification}% Force line breaks with \\

\author{Marco Leonetti}
\email{marco.leonetti@cnr.it}
\affiliation{Soft and Living Matter Laboratory, Institute of Nanotechnology, Consiglio Nazionale delle Ricerche, 00185 Rome, Italy}
\affiliation{Center for Life Nano- \& Neuro-Science, Italian Institute of Technology, Rome,Italy}
\affiliation{Rebel Dynamics-IIT CLN2S Jointlab, 00161 Roma Italy}

\author{Giorgio Gosti}%
\affiliation{ Soft and Living Matter Laboratory, Institute of Nanotechnology, Consiglio Nazionale delle Ricerche, 00185 Rome, Italy}
\affiliation{Center for Life Nano- \& Neuro-Science, Italian Institute of Technology, Rome,Italy}

\author{Giancarlo Ruocco}%
\affiliation{Center for Life Nano- \& Neuro-Science, Italian Institute of Technology, Rome,Italy}
\affiliation{Department of Physics, University Sapienza, I-00185 Roma, Italy}

%\affiliation{
%\affiliation{
%\affiliation{%

%\collaboration{}%\noaffiliation

\date{\today}% It is always \today, today,

\begin{abstract}
Disorder is a pervasive characteristic of natural systems, offering a wealth of non-repeating patterns. In this study, we present a novel storage method that harnesses naturally-occurring random structures to store an arbitrary pattern in a memory device. This method, the stochastic emergent storage (SES), builds upon the concept of emergent archetypes, where a training set of imperfect examples (prototypes) is employed to instantiate an archetype in an Hopfield-like network through emergent processes.

We demostrate  this non-Hebbian paradigm in the photonic domain by utilizing random transmission matrices, which govern light scattering in a white-paint turbid medium, as prototypes. Through the implementation of programmable hardware, we successfully realize and experimentally validate the capability to store an arbitrary archetype and perform classification at the speed of light. Leveraging the vast number of modes excited by mesoscopic diffusion, our approach enables the simultaneous storage of thousands of memories without requiring any additional fabrication efforts. Similar to a content addressable memory, all stored memories can be collectively assessed against a given pattern to identify the matching element. Furthermore, by organizing memories spatially into distinct classes, they become features within a higher-level categorical (deeper) optical classification layer.

\end{abstract}

\maketitle

%\tableofcontents
Neural networks have made significant contributions to the field of Artificial Intelligence, serving as both a tool for mathematical modeling and a means to understand brain function. The Hopfield paradigm \cite{hopfield1982neural,amit1985storing} has played a crucial role in this domain, utilizing a synaptic matrix to represent the interconnections between neurons. This matrix possesses the remarkable ability to store and recognize patterns, and serve as a fundamental framework for the realization of future content-addressable memory (CAM) \cite{farhat1985optical,quashef2022ultracompact}.

To store a memory consisting of N elements, the widely adopted approach is to employ Hebb's rule \cite{hebb19680}. This rule entails constructing a synaptic matrix, denoted as $\bm{T}$, by taking the tensorial product of the vector $\bm{\phi^*}$ (representing the pattern to be stored) and its conjugate transpose ($\bm{\phi}^{\bm{*}\dag}$):

\begin{eqnarray}
\bm{T}=\bm{\phi^*}\otimes \bm{\phi}^{\bm{*}\dag}
\end{eqnarray}

However there is a fundamental limit to the number of memories that can be reliably stored using Hebbian-based approaches. As the network becomes more densely populated, the interactions between different memory elements can lead to the emergence of unintended and uncontrolled memory states \cite{amit1985storing}.
To address this limitation, recent research has explored various methods to enhance the  capacity of neural networks: dilution \cite{Brunel2016,Folli2018,Mocanu2018,
Leonetti2020}, autapses \cite{Folli2017,Gosti2019}, and convex probability flow \cite{Hillar2018,Hillar2021}.

It has been recently proposed to use the interaction among stored patterns in a constructive way:  an \textit{emergent archetype} may be stored by proposing to the network multiple \textit{prototypes} that closely resemble the target pattern but are intentionally corrupted or filled with errors. The interaction between these prototypes serves to strengthen the emergence of the desired memory \cite{agliari2022emergence}.
This paradigm is connected to the prototype concept developed in hierarchical clustering, in which prototypes are elements of the dataset representative of each cluster \cite{Bien2012}

\begin{figure*}[th!]
\includegraphics[width=\textwidth]{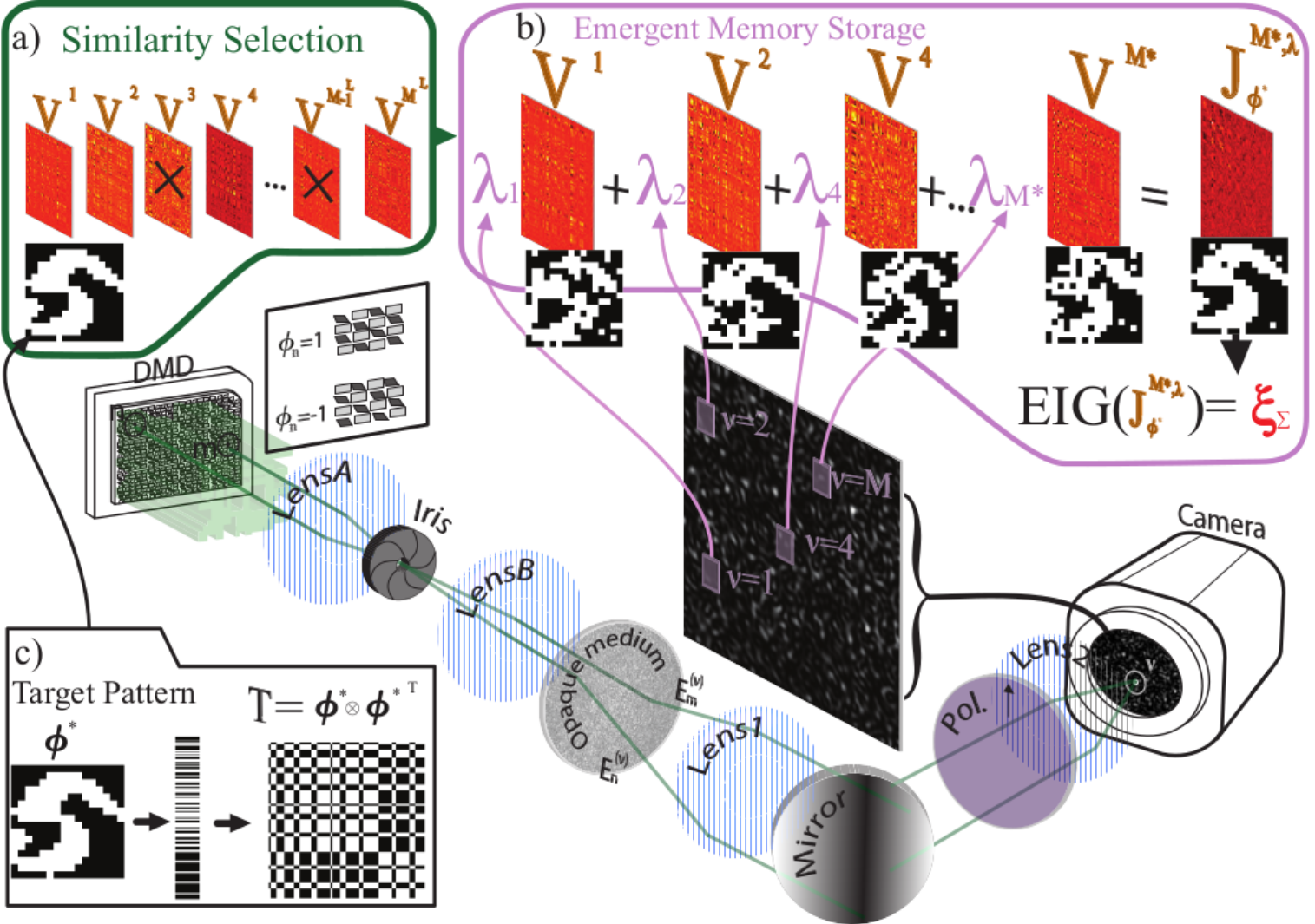}
\caption{\textbf{Emergent memory storage scheme.} The sketch represents the experimental setup employed both to present the input query $\bm{\phi}$ and to measure optical synaptic matrix $\bm{V^\nu}$ (\textbf{details in the methods}). Note that the magnification is engineered
in order %such as
to have a mode for each camera pixel ($\nu$ is the pixel /mode index).  Panel a) shows the similarity selection stage.
Panel b) shows the reconstruction process for %the process to construct of
the emergent memory $\bm{\xi}_\Sigma$ with the optical operator $\bm{J}^{\bm{\mathcal{M}}^{*},\bm{\lambda}}_{\bm{\phi^*}}$. Panel c) shows a sample pattern and %how
the relative synaptic matrix constructed with the Hebb's rule. }
\label{Setup}
\end{figure*}

In this study, we introduce a novel learning strategy called \textit{stochastic Emergent Storage} (SES). SES  taps into the abundance of natural randomness to construct an emergent representation of the desired memory. Capitalizing a vast database of fully random patterns freely produced by a disordered, self-assembled structure, we select a set of prototypes that bear resemblance to the target memory through a similarity-based criterion. Subsequently, by performing a weighted sum of the synaptic matrices corresponding to these selected prototypes, we are able to effectively generate the desired pattern in an emergent fashion.

Given the inherent advantages of photonic computation, such as ultra-fast wavefront transformation and parallel operation,
it results that optics is the ideal domain
to explore the SES paradigm. The convergence of photonics, artificial intelligence, and machine learning represents a highly active and promising area of research \cite{farhat1985optical,fu2023photonic}, leading to  novel interdisciplinary paradigms such as  Diffractive Deep neural networks \cite{lin2018all,liu2022programmable} photonic Ising machines \cite{pierangeli2019large} and photonic  Boltzmann computing Machines~\cite{yamashita2023spatial}.
However, these approaches typically  rely on direct control over optical properties of millions of scattering elements, which can be challenging and costly both with microfabrication or adaptive optical elements.

In a departure from traditional approaches, disordered scattering structures have been proposed as a radically different avenue for optical computation in various applications: classification \cite{saade2016random}, vector-matrix multiplication \cite{ohana2023linear}, computation of statistical mechanics ensembles dynamics \cite{leonetti2021optical}, and others \cite{brossollet2021lighton}.

Here, we propose to employ the scattering intrinsic patterns, the optical transmission matrices,  to realize a SES-based optical hardware, the \textit{disordered classifier}. This device is  capable of efficiently performing pattern storage, and subsequent pattern retrieval.
It is able
to simultaneously compare an input pattern with
thousands of stored elements, and it enables a two-layer architecture, providing categorical (deep) classification, which allows
for more complex tasks.

The idea stems from the fact that intensity scattered by a disordered medium
into a mode $\nu$ resulting from an input %test
pattern $\bm{\phi}$
%by a disordered medium
may be written as:

\begin{eqnarray}
I^\nu(\bm{\phi}) =\bm{\phi}\cdot\bm{V^\nu}\cdot\bm{\phi}^\dag
\end{eqnarray}
with the scattering process driven by the matrix:
\begin{eqnarray}
\bm{V}^\nu =\bm{\xi}^\nu\otimes \bm{\xi}^{\nu \dag}
\end{eqnarray}
generated from the tensorial product of the transmission matrix row (transmission vector) $\bm{\xi}^{\nu }$ with its conjugate transpose $\bm{\xi}^{\nu \dag}$.

Indeed  $I^\nu(\bm{\phi})$ is maximized if $\bm{\phi} \parallel \bm{\xi}^\nu$: this paradigm is at the basis of the wavefront shaping techniques  \cite{vellekoop2007focusing,vellekoop2010exploiting}, in which the input pattern is adapted to the transmission matrix elements. Thus, scattering into a mode (corresponding to one of our camera pixels, see \textbf{methods}) is described by the same mathematics of the
Hopfield Hamiltonian
%``single pass'' Hopfield dynamics, (without threshold)
and a pattern  is ``recognized'' (produces maximal intensity) if it matches the $\bm{\xi}^\nu$ vector. Given this mapping, $\bm{V^\nu}$ may be named an \textit{optical synaptic matrix}.

In naturally  occurring scattering, one has no control over the pattern $\bm{\xi}^\nu$
and the relative optical synaptic matrix $\bm{V}^\nu$
because it results from a multitude of subsequent scattering events with micro-nano particles of unknown shape, optical properties, and location.
Here we propose to store an arbitrary, user-defined, memory (or pattern)
in naturally occurring scattering media, by exploiting the fact that a scattering process generated billions of output modes,  each with a unique and random embedded memory pattern $\bm{\xi}^\nu$
and the relative $\bm{V}^\nu$.
%We will exploit a linear combination of them
Thus we propose a new method to realize a photonic linear combination of  $\bm{V}^\nu$ to generate an arbitrary optical synaptic matrix.

First, we will employ this  to  realizing an optical equivalent of the Hebbs rule:  the \textit{stochastic  Hebbs storage} (SHS).
Then we will show how the storage and classification performance is greatly improved if SES is exploited.

With the SHS  we want to generate a synaptic optical matrix $\bm{J}^{\bm{\mathcal{M}},\bm\lambda}_{\bm{T}}$ equivalent to an Hebb's matrix $\bm{T}$ with the aim  to store the pattern $\bm{\phi}^*$. To do this, we rely on a  linear combination of a set
$\bm{\mathcal{M}}=\{\bm{V}^1, \bm{V}^1 \ldots \bm{V}^M\}$
of random optical synaptic matrices resulting from uncontrolled scattering:

\begin{eqnarray}
\bm{J}^{\bm{\mathcal{M}},\bm{\lambda}}_{\bm{T}}=\sum^{M}_\nu \lambda^\nu \bm{V}^\nu
\end{eqnarray}
Indeed the matrix $\bm{J}^{\bm{\mathcal{M}},\bm\lambda}_{\bm{T}}$ is connected to the intensities of the modes pertaining to the set  $\bm{\mathcal{M}}$  with the following equation:

\begin{eqnarray}
\bm{\phi}\cdot\bm{J}^{\bm{\mathcal{M}},\bm{\lambda}}_{\bm{T}}\cdot\bm{\phi}^\dag= \sum^M_\nu \lambda^\nu I^\nu(\bm{\phi})=I^{\bm{\mathcal{M}},\bm{\lambda}}_{\bm{T}}(\bm{\phi}).
\label{Transformed Intensity SHS}
\end{eqnarray}
for which we name  $I^{\bm{\mathcal{M}},\bm{\lambda}}_{\bm{T}}(\bm{\phi})$  the \textit{transformed intensity} generated by the optical operator $\bm{J}^{\bm{\mathcal{M}},\bm{\lambda}}_{\bm{T}}$ emulating the Hebb's synaptic matrix ${\bm{T}}$.

\begin{figure}
\includegraphics[width=8 cm]{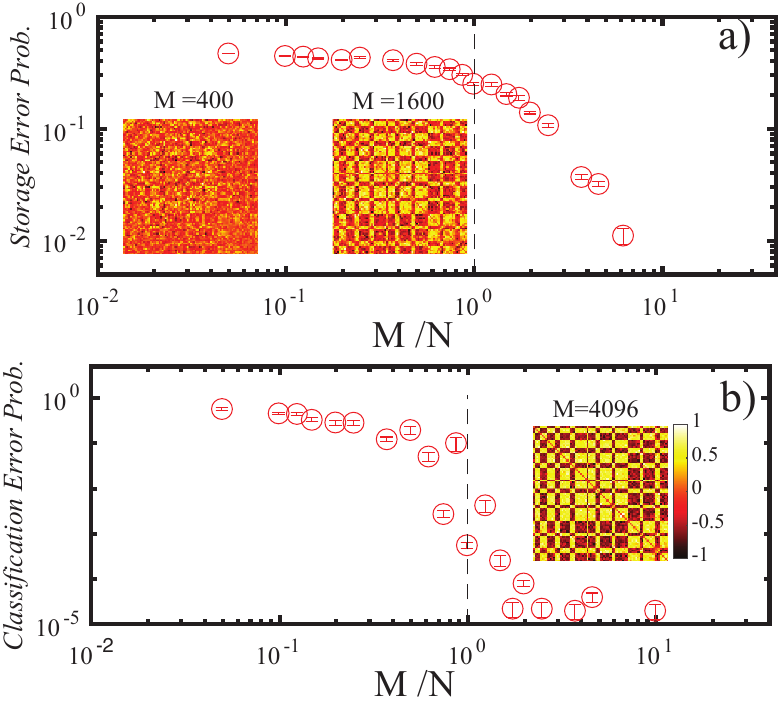}
\caption{ \textbf{Photonic Stochastic Hebb's Storage}. In a) the \textit{storage error probability} with respect to  $M/N$. In b) the \textit{Classification error probability} versus  $M/N$. The inserts report the images of the reconstructed  $\bm{J}^{\bm{\mathcal{M}},\bm{\lambda}}_{\bm{T}}$ matrix for various values of $M$. Note that the target $\bm{T}$ is shown in Fig \ref{Setup}c. $N=81$}
\label{Hopfield}
\end{figure}

The values for coefficients  $\lambda^\nu$ are obtained by a Monte Carlo algorithm, \textbf{(see methods)} minimizing  the difference between the target matrix and $\bm{J}^{\bm{\mathcal{M}},\bm{\lambda}}_{\bm{T}}$. Each coefficient may be then realized in hardware  (mode-specific neutral density filters) or software fashion.

Employing SHS we can design any arbitrary  optical operator if the two following ingredients are available: i) the access to the intensity $I^\nu(\bm{\phi})$ produced by a sufficiently large number of modes and ii) the correspondent optical synaptic matrix  $\bm{V}^\nu$ for each mode. This is now possible with the Complete Couplings Mapping Method (CCMM, \textbf{see methods}), which enables the measurement of the intrinsic (no interference with a reference) $\bm{V}^\nu$ with a Digital Micromirror Device (DMD).

With the CCMM, and the experimental apparatus shown in Fig. \ref{Setup} and detailed also in  the \textbf{methods}  we are able to gather a repository $\bm{\mathcal{M}}^L$ of tens of thousands ($M^L= 65536$) of optical synaptic matrices in minutes from which we sample
a random subset $\bm{\mathcal{M}}$ (with $M$ random samples) which we use
as bases to construct our target synaptic matrix.

The performance of this optical learning approach is shown in Fig. \ref{Hopfield}, in which we realized an Hebb's dyadic-like optical synaptic matrix  (see insets of Fig. \ref{Hopfield} ) from a ZnO scattering layer.

The memory pattern stored in our system is
$\bm{\xi_\Sigma}=SEIG(\bm{J}^{\bm{\mathcal{M}},\bm{\lambda}}_{\bm{T}})$, with $SEIG(\bm{H})$ the operator that finds the  eigenvector correspondent to the largest eigenvalue of $\bm{H}$
and then produces a binary vector with its elements' sign. Performance in storage and classification for SHS  are reported respectively in \ref{Hopfield}a and \ref{Hopfield}b \textbf{(see methods)}. SHS is \textit{basis hungry}, requiring a large number of random optical synaptic matrices (which means modes/sensors/pixels) to successfully  construct a memory element.

For the remainder of the paper, we will discuss how the performance  drastically improves with SES. We  recognize that  each optical synaptic matrix contains a memory  $\bm{\xi}^\nu=SEIG(\bm{V}^\nu)$ then (instead of randomly extracting modes) we perform a similarity selection (see Fig. \ref{Setup}a and \textbf{methods}) in which we extract a set  ${M}^*$  whose intrinsic memories are the closest possible to the target pattern $\bm{\phi^*}$ (see insets of Fig. \ref{Setup}).
The fact that in a mesoscopic laser scattering process, billions of independent modes can be produced and millions of them can be measured at once with modern cameras, is strategically employed in SES to boost the performance.

These selected $\bm{\xi}^\nu$ can be seen as prototypes of the target archetype, i.e. imperfect representations of the pattern to be stored (such as the one in Fig. \ref{Setup}c).  In SES, these prototypes interact constructively, generating a representation of the memory $\bm{\phi^*}$ in an emergent fashion \cite{agliari2022emergence}. The attenuation coefficients $\bm{\lambda}$ are found  by minimizing the distance between the archetype pattern to be stored ${\bm{\phi^*}}$ and the matrix first eigenvector $SEIG(\bm{J}^{\bm{\mathcal{M}}^{*},\bm{\lambda}}_{\bm{\phi^*}})=
\bm{\xi_\Sigma}$  (see \textbf{methods}).
The transformed intensity in SES reads :
\begin{eqnarray}
I^{\bm{\mathcal{M}}^{*},\bm{\lambda}}_{\bm{\phi^*}}(\bm{\phi})=\bm{\phi}\cdot\bm{J}^{\bm{\mathcal{M}}^{*},\bm{\lambda}}_{\bm{\phi^*}}\cdot\bm{\phi}^\dag= \sum^{M^{*}}_\nu \lambda^\nu I^\nu(\bm{\phi}).
\label{Transformed Intensity SES}
\end{eqnarray}

\begin{figure}
\includegraphics[width=8 cm]{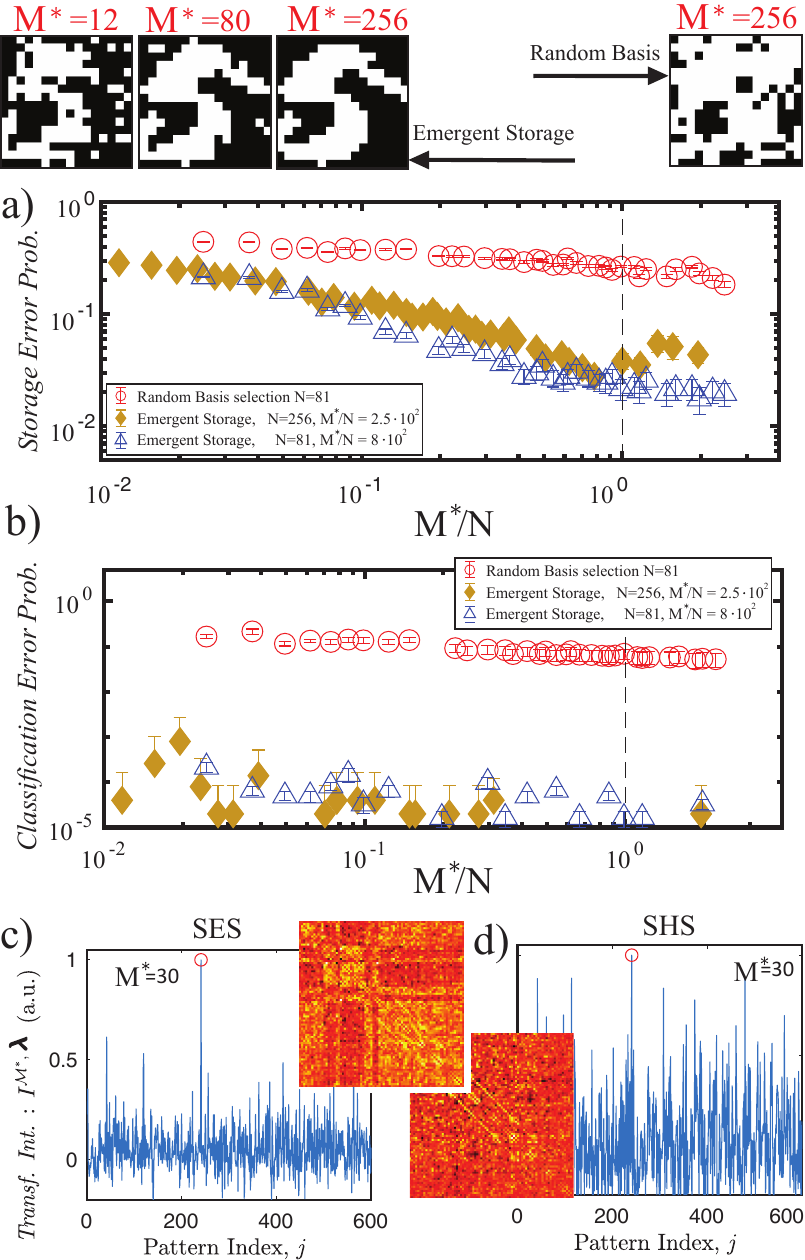}
\caption{\textbf{Photonic Stocastic Emergent storage}. Top panels show the stored patterns obtained with SES
for different values of $M^*$ with
Similarity selection (three patterns on the left),
and with random selection (pattern on the right).
Panel a) shows \textit{storage error probability} while panel   b) the \textit{Classification error probability}. Both  with respect to  $M^*/N$. c) and d): Transformed intensity for 600 patterns in the repository (pattern index $j$ on the ordinate axis) for the SES and SHS ($j=241$, red circled intensity, correspond to the stored pattern. The insets between panels c) and d) report the obtained $\bm{J}^{\bm{\mathcal{M^*}},\bm{\lambda}}_{\bm{\phi^*}}$}
\label{Emergent Learning}
\end{figure}

The potential of SES is clarified in  Fig. \ref{Emergent Learning}: the panels on the top left represent the stored pattern (target pattern is reported in \ref{Setup}c)
for various levels %values
of $M^*$. Note that SES greatly outperforms the random selection approach where emergent storage is absent (panel on the right).

Panel \ref{Emergent Learning}a shows the storage capacity of the system. Blue triangles are relative to patterns with ${N}=81$ elements, while for golden diamonds ${N}=256$. The Storage error probability (the lower the better, \textbf{see methods}) improves  more than an order of magnitude with respect to random selection (red circles). Classification error (the lower the better, panel \ref{Emergent Learning}b, \textbf{see methods}) is three to four orders of magnitude better with respect to the randomly selected database. Note that the SES enormously outperforms SHS, indeed
it is %is it %Era posto comem domanda.
possible to perform classification in the $M<< N$ configuration, i.e. employing a number of camera pixels($M$) much smaller than the elements composing the pattern $N$.

Panel \ref{Emergent Learning}c and  \ref{Emergent Learning}d, show
a classification process example. %an example of classification process.
The emergent learning process has been employed to store the pattern $\bm{\phi}$ with index  $j=241$  from a repository of 5000 patterns. Panel  \ref{Emergent Learning}c reports \textit{transformed intensity} for the first 600 repository elements:
a clear peak
is distinguishable
at $j=241$
this implies that the pattern is recognized).
The same graph is shown for the case of the random basis case, in which classification is more noisy.

Our \textit{disordered  classifier} can work in parallel, simultaneously comparing an input with all memories stored, effectively working as a content addressable memory. \cite{quashef2022ultracompact}.
%TODO: qui potrebbe essere dove si parla dei pattern nelle rei conv...

\begin{figure}[ht!]
\includegraphics[width=8cm]{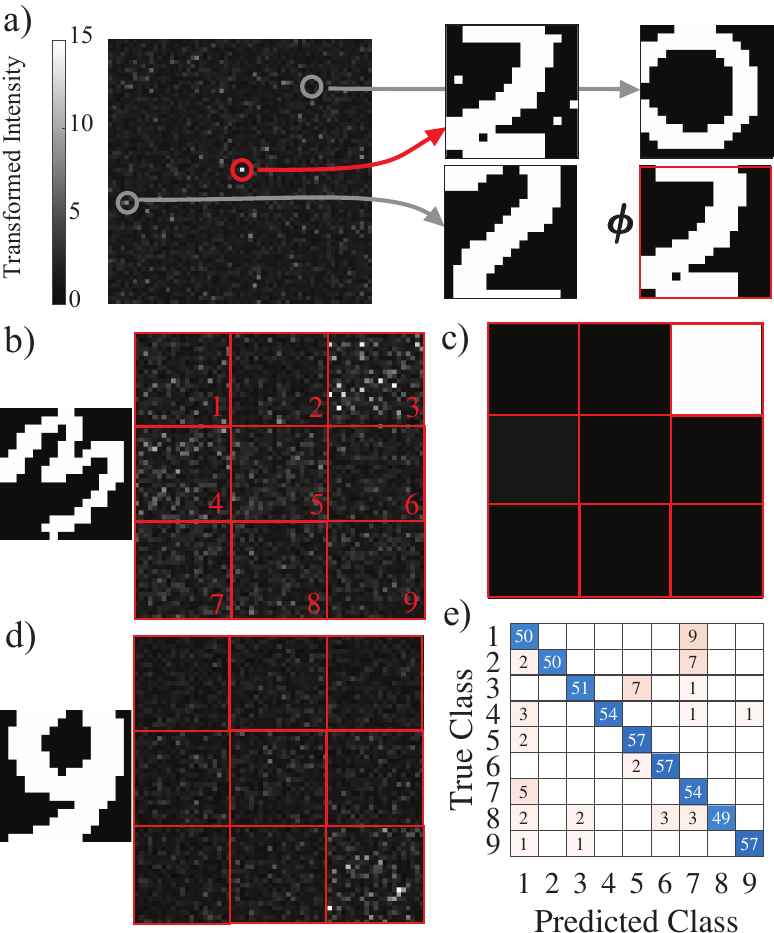}
\caption{\textbf{Parallel/Categorical  photonic classification.} Panel a) Shows the \textit{transformed intensities} relative to 4096 stored memories ($N=256$, $M=120$) organized in 64 $\times$ 64 camera-like diagram.  The \textit{transformed intensities} are all generated as the pattern $\bm\phi$ (shown on the right, highlighted by a red frame)  is presented to the \textit{disordered classifier}. For the circled pixels, the correspondent stored memory patterns are shown on the right (indicated by the arrow). The red circled pixel is associated to the memory which is the most similar to the pattern presented with the DMD. The diagram in b) is similar to the previous but memories associated with 9 numerical categories, are organized in quadrants. The presented pattern $\bm{\phi}$ is the ``three'' on the left).  Panel c) shows the thresholded and integrated intensity corresponding to the measure in b). Panel d) is the same as b) but with a ``nine'' pattern presented as input. Panel d) reports the confusion matrix for the categorical (number class) recognition (531 patterns in the validation set, 90.21\% recognition efficiency).  }
\label{Higher rank category}
\end{figure}

The experimentally retrieved \textit{transformed intensity} for 4096 different memory elements $\bm{\phi^*}$ is reported in Fig. \ref{Higher rank category}  ( organized in a camera-like 64 $\times$ 64 pixels diagram) for the proposed pattern $\bm{\phi}$. Each value of  $I^{\bm{\mathcal{M}}^{*},\bm{\lambda}}_{\bm{\phi^*}} (\bm{\phi})$  represent the degree of similitude of $\bm{\phi}$ to $\bm{\phi^*}$.  The patterns to the right side of the panel report the proposed pattern $\bm{\phi}$ and the stored patterns relative to each arrow-indicated pixel. The pixel indicated with a red circle contains the pattern most similar to $\bm{\phi}$ thus as expected produces the highest intensity. The system effectively works as a CAM in which an input query $\bm{\phi}$ is tested in parallel against a list of stored patterns (the $\bm{\phi}^*$) identifying the matching memory as the most intense \textit{ transformed intensity} pixel.

The interplay between Hopfield networks and Deep learning has been recently proposed and investigated \cite{ramsauer2020hopfield,krotov2023new}. In this framework here we demonstrate a new approach to perform higher rank categorical classification employing the cashed memories as features \cite{molnar2020interpretable,zeiler2014visualizing}.
We tested it on a 4500 randomly tilted digits images repository which is organized into 9 categories (digits from 1 to 9). We stored 3969 patterns/features in the \textit{disordered classifier}
(441 per each digit), %(441 per each digit),
leaving 59 patterns per category for validation. In the camera-like diagrams (Fig. \ref{Higher rank category}c and \ref{Higher rank category}e ) each category is found in the correspondent quadrant of the image. The two panels show the response of the \textit{disordered classifier} to the inputs on the left for which the correspondent quadrants show an high number of intense pixels. Panel \ref{Higher rank category}d show integrated intensity after thresholding. Figure  \ref{Higher rank category}f reports the confusion matrix for all labels, demonstrating categorical recognition efficiency higher than 90\% .

In summary, the stochastic Emergent Storage (SES) paradigm enables classification with a significantly smaller number of sensors/pixels/modes compared to the elements composing the pattern. This opens up the possibility of fabricating complex pattern classifiers with only a few detecting elements, eliminating the fabrication processes.
SES offers an disruptive paradigm for pattern learning
and network training, a paradigm that
%The potential of the SES paradigm
can be potentially
applied to any other disordered systems, such as biological neural networks or neuromorphic computer architectures.
Exploring the emergent learning process in these systems can also provide valuable insights into the memory formation process in the brain.

The results presented in this study contribute to the ongoing challenge of understanding the biological memory formation process. The connectionist hypothesis \cite{hebb19680}, which suggests that neural networks form new links or adjust existing ones when storing new patterns, and the innate hypothesis\cite{Perin2011}, which posits that patterns are stored using pre-existing neural assemblies with fixed connectivity, are still subjects of debate.
SES introduces a fresh perspective to the problem by leveraging the Hebbian structure of the synaptic matrix, with a foundation of the connectionist hypothesis. However, SES goes beyond by exploring the potential of a stochastic \textit{innate} network in which,
pre-determined random synaptic structures are combined to generate memory elements in an emergent manner.
\\

\begin{acknowledgments}
This research was funded by:  Regione Lazio, Project LOCALSCENT, Grant PROT. A0375-2020-36549, Call POR-FESR ``Gruppi di Ricerca 2020'' (to M.L.); ERC-2019-Synergy Grant (ASTRA, n. 855923; to GR); EIC-2022-PathfinderOpen (ivBM-4PAP, n. 101098989; to G.R.); Project ``National Center for Gene Therapy and Drugs based on RNA Technology'' (CN00000041) financed by NextGeneration EU PNRR MUR - M4C2 – Action 1.4 - Call ``Potenziamento strutture di ricerca e creazione di campioni nazionali di R\&S''  (CUP J33C22001130001) (to G.R.). The authors Acknowledge Enrico Ventura, and Luigi Loreti for fruitful discussions.
\end{acknowledgments}

\begin{acknowledgments}
The authors declare no competing interests
\end{acknowledgments}

\bibliography{Main_BIB}

\clearpage

\section{Methods }
\subsection{Background}
In our experiment (similarly to a typical wavefront shaping experiment), light from a coherent source is controlled by a spatial light modulator and transmitted after propagation through a disordered medium into the mode $\nu$. The field transmitted  at $\nu$ is described as

 \begin{equation}
E^{\nu}=\sum _{n=1}^N E^\nu_n \phi_n %E^{\nu}=\sum _{n=1}^\mathcal{N} E^\nu_n \phi_n
\end{equation}

where the index $n$ runs on the controlled segments at the input of the disordered medium, $E^\nu_n$ is the field resulting from laser field from the  $n th$ segment transformed by the transmission matrix element on the sensor $\nu$ and $\phi^\nu_n$ is the phase value from the wavefront modulator. In our experiment, we consider the simplified configuration in which $\phi^\nu_n \in \{-1, +1\}$.

The field at $\nu$ can be separated in its two components: the field-at-the-segment $A_n$ and transmission matrix elment $t^\nu_n$
 \begin{equation}
E^\nu_n =A_n t^\nu_n
\end{equation}

Indeed the $E^\nu_n$ are Gaussianly distributed complex numbers:
 \begin{equation}
E^\nu_n = \xi^\nu_n +i \eta ^\nu_n.
\end{equation}

In the case
in which
just two segments $n$ and $m$ are active and in the $+1$ configuration, we can ignore the $\phi_n$:

\begin{equation}
E^\nu = E^\nu_n +E^\nu_m= \xi^\nu_n +i \eta ^\nu_n +\xi^\nu_m +i \eta ^\nu_m .
\end{equation}

%Intensity in absence of modulation
In absence of modulation intensity
is written as the modulus square of $E^\nu$ %the previous

 \begin{eqnarray}
I^\nu =& |E^\nu_n +E^\nu_m||E^{\nu \dag}_n +E^{\nu\dag}_m| = \nonumber \\
&|E^\nu_n|^2+ |E^\nu_m|^2 +|E^\nu_n||E^{\nu \dag}_m|+|E^{\nu \dag}_n||E^{\nu}_m|.
 \end{eqnarray}

we recognize that
 \begin{eqnarray}
|E^\nu_n||E^{\nu \dag}_m|=\xi^\nu_n\xi^\nu_m   -i\xi^\nu_n\eta ^\nu_m +i \eta ^\nu_n\xi^\nu_m  + \eta ^\nu_n\eta ^\nu_m \\
|E^{\nu \dag}_n||E^{\nu}_m|=\xi^\nu_n\xi^\nu_m   +i\xi^\nu_n\eta ^\nu_m -i \eta ^\nu_n\xi^\nu_m  + \eta ^\nu_n\eta ^\nu_m \\
|E^\nu_n||E^{\nu \dag}_m|+|E^{\nu \dag}_n||E^{\nu}_m|=2\xi^\nu_n\xi^\nu_m+2\eta ^\nu_n\eta ^\nu_m.
 \end{eqnarray}

 thus

 \begin{eqnarray}
I^\nu = \xi^{\nu 2}_n+\eta ^{\nu 2}_n+\xi^{\nu 2}_m+\eta ^{\nu 2}_m+2\xi^\nu_n\xi^\nu_m+2\eta ^\nu_n\eta ^\nu_m .
 \end{eqnarray}
or
 \begin{eqnarray}
I^\nu = E^{\nu 2}_n+E^{\nu 2}_m+2\xi^\nu_n\xi^\nu_m+2\eta ^\nu_n\eta ^\nu_m .
\label{twomirror}
 \end{eqnarray}

in general for $N$ segments in an arbitrary configuration % $\mathcal{N}$
 \begin{eqnarray}
I^\nu = \sum_{n,m}^{N} E^{\nu 2}_n+E^{\nu 2}_m+2(\xi^\nu_n\xi^\nu_m+\eta ^\nu_n\eta ^\nu_m)\phi_n\phi_m.
%I^\nu = \sum_{n,m}^{\mathcal{N},\mathcal{N}} E^{\nu 2}_n+E^{\nu 2}_m+2\xi^\nu_n\xi^\nu_m+2\eta ^\nu_n\eta ^\nu_m .
 \end{eqnarray}

the argument of the sum can be written in matrix form defining the matrix $\bm{V}^\nu$ also named optical coupling matrix:

 \begin{eqnarray}
V^\nu_{nn} = E^{\nu 2}_n=\xi^{\nu 2}_n+\eta ^{\nu 2}_n. \\
V^\nu_{nm} = +\xi^\nu_n\xi^\nu_m+\eta ^\nu_n\eta ^\nu_m
 \end{eqnarray}
Matrix $\mathbf{V^\nu}$ is a bi-dyadic matrix and it can be rewritten in matricial notation:
\begin{eqnarray}
\mathbf{V^\nu}=  \bm{\xi}^\nu\otimes\bm{\xi}^{\nu \dag}+\bm{\eta}^\nu\otimes\bm{\eta}^{\nu \dag}
 \end{eqnarray}

where the  $\bm{notation}$  indicates a vector on lowercase Greek letters and a matrix on uppercase, while  $^\dag$ is the conjugate transpose operator.
Being bi-dyadic the matrix possesses the eigenvectors
$\bm{\xi^\nu}$ and  $
\bm{\eta^\nu}$ by construction.

When modulation is present with an input modulation pattern $\bm{\phi}$

\begin{eqnarray}
I^\nu(\bm{\phi}) =\sum_{n,m}^{N} V^\nu_{nm}\phi_n\phi_m=\bm{\phi}\cdot\bm{V^\nu}\cdot\bm{\phi}^\dag
\end{eqnarray}

The optical operator $\bm{V}^{\nu}$, associates thus the  pattern/array $\bm{\phi}$ to the scalar $I^{\nu}$ which is a measure of the degree of similitude of $\bm{\phi}$ to the first eigenvector of $\bm{V}^{\nu}$, $EIG(\bm{V}^{\nu})= \bm{\xi^\nu}$.

Note that to simplify the realization of the experiment, we operate in the configuration in which each mode $\nu$ corresponds to a single sensor. As we employ a camera to measure $I^\nu$,
%the
e the one-mode-per-pixel configuration is obtained  by properly tuning the optical magnification.

\subsection{Stochastic Hebb's storage protocols details}

By summing intensity measured at two modes $\nu_1$ and $\nu_2$, and considering linearity of the process:
\begin{eqnarray}
I^{\nu_1}+I^{\nu_2}&= \bm{\phi} \cdot\bm{V}^{\nu_1}\cdot\bm{\phi}^\dag+\bm{\phi} \cdot\bm{V}^{\nu_2}\cdot\bm{\phi}^\dag=\\
& =\bm{\phi} \cdot\bm{J}^{\nu_1,\nu_2}\cdot\bm{\phi}^\dag. \nonumber
\end{eqnarray}
Generalizing, i.e. summing intensity at an arbitrary number ${M}$ of modes pertaining to the set $\bm{\mathcal{M}}=\{\nu_1, \nu_2, ... \nu_M \}$, we retrieve

\begin{eqnarray}
I^{\bm{\mathcal{M}}}=\sum_{\nu}^{M}\bm{\phi}  \cdot\bm{V}^{\nu}\cdot\bm{\phi}^\dag=\bm{\phi}  \cdot\bm{J}^{\bm{\mathcal{M}}}\cdot\bm{\phi}^\dag
%\sum_{n,m}^{N} \sum_{\nu}^M {V^{\nu}_{nm}}=\sum_{n,m}^{N} {J_{nm}}
%I^\mathcal{M}=\sum_{n,m}^{\mathcal{N},\mathcal{N}} \sum_{\nu}^\mathcal{M} {V^{\nu}_{nm}}=\sum_{n,m}^{\mathcal{N},\mathcal{N}} {J_{nm}}
\end{eqnarray}

with
\begin{eqnarray}
J^{\bm{\mathcal{M}}}_{nm}=\sum^{{{M}}}_{\nu}{V^{\nu}_{nm}}
\end{eqnarray}

Thus,  the optical operator $\bm{J}^{\bm{\mathcal{M}}}$ associates %thus
a pattern/array $\bm{\phi}$ to the scalar $I^{\bm{\mathcal{M}}}$, the \textit{transformed  intensity}, which is a measure of the degree of similitude of $\bm{\phi}$ to the first eigenvector of $\bm{J}^{\bm{\mathcal{M}}}$: $EIG(\bm{J}^{\bm{\mathcal{M}}})= \bm{\xi}_{\bm{\mathcal{M}}}$.
To deliver an user-designed arbitrary optical operator, we introduce the tailored attenuation coefficients $\lambda^\nu \in [ 0, 1]$.
These can be both obtained in ``software version'' (multiplying each $I^\nu$ by an attenuation coefficient $\lambda^\nu$) or by realizing a mode-specific hardware optical attenuator (such as proposed in the sketch in Fig. 1, fuchsia windows, \textbf{see later section}).

\textit{Transformed intensity} with the addition of the attenuation coefficients reads as:

\begin{eqnarray}
I^{{\bm{\mathcal{M}}},\bm{\lambda}}=\sum_{\nu}^{M}\bm{\phi}  \cdot \lambda^\nu \bm{V}^{\nu}\cdot\bm{\phi}^\dag=\bm{\phi}  \cdot\bm{J}^{\bm{\mathcal{M}} ,\bm{\lambda} }\cdot\bm{\phi}^\dag
\label{SHS tranf int}
\end{eqnarray}

In SHS, the  absorption coefficients  $\bm{\lambda}$ are the free parameters which enable to design the arbitrary optical operator $\bm{J}^{\bm{\mathcal{M}} ,\bm{\lambda} }$. For example, to replicate the dyadic matrix constructed with he Hebb's rule $\bm{T}$ and capable to store the pattern $\bm{\phi}$ (see Fig 1c of the main paper) one has to select $\bm{\lambda}$ so that the function

\begin{eqnarray}
\mathcal{F}({\bm{\mathcal{M}}},\bm\lambda)=\sum_{n,m}^{N} \left| \sum_{\nu}^M \lambda^{\nu} V^{\nu}_{nm} - T_{nm}\right|^2= \nonumber \\
=DIST\left(\bm{J}^{M,\bm{\lambda}},  \bm{T}\right)
\label{SHS min fun}
\end{eqnarray}
is minimized. We name $\bm{J}_{\bm{T}}^{\bm{M,\lambda}}$ the optical synaptic matrix in which $\bm{\lambda}$ have been optimized to deliver the optical operator $\bm{T}$, and
\begin{eqnarray}
I_{\bm{T}}^{\bm{M,\lambda}}=\sum_{\nu}^{M} \lambda^\nu I^{\nu}
\label{trained transformed intensity}
\end{eqnarray}

the relative \textit{transformed intensity}.

This approach employs the random, naturally-occurring optical synaptic matrices from the set ${\bm{\mathcal{M}}}$ as a random basis on which to build the target optical operator. Its effectiveness is thus dependent on the number of free parameters with respect to the constraints. The constraints are the number of independent elements that have to be tailored on ${\bm{{T}}}$. These are $\Pi=(N(N-1)/2$ as ${\bm{{T}}}$ is symmetric. Indeed as shown in Fig. 2 of the main paper (inset of panel b) for the $N=81$ case, it is possible to replicate almost identically ${\bm{{T}}}$  when $M>\Pi$, that is when the number of free parameters (the $\bm{\lambda}$)  is comparable with the constraints.

\subsection{Storage Error Probability}
In our storage paradigm, the stored pattern corresponds to the eigenvector of the ${\bm{{T}}}$. As we are employing binary patterns,
the sign operation is needed. The stored pattern is thus $SEIG(\bm{J}^{\bm{\mathcal{M}}^{*},\bm{\lambda}}_{\bm{\phi^*}})= \bm{\xi_\Sigma}$, where the $SEIG()$ operator retrieves the first eigenvalue of a matrix and applies the sign operation to it. The \textit{Storage Error Probability} reported in Fig.s 2 and 3, represent the a measure of the storage process effectiveness. First, we calculate the number of elements of  $\bm{\xi_\Sigma}$ which differ from the target memory $\bm{\phi^*}$, $S\_ERR$. The value of $S\_ERR$  can be seen as  the number of error pixels in the stored pattern image.

Then we compute
\begin{eqnarray}
 \text{\textit{Storage Error Probability} =} S\_ERR/N.
\end{eqnarray}

For storage purposes,  obviously the lower, the \textit{Storage Error Probability} the better.
\newline

\subsection{Classification Error Probability}
The optical operator $\bm{J}_{\bm{T}}^{\bm{M,\lambda}}$ associates the \textit{ transformed intensity } scalar to each input pattern $\bm{\phi}$:

\begin{eqnarray}
I_{\bm{T}}^{{\bm{\mathcal{M}}},\bm{\lambda}}=\bm{\phi}  \cdot\bm{J}_{\bm{T}}^{\bm{\mathcal{M}} ,\bm{\lambda} }\cdot\bm{\phi}^\dag.
\label{SHS tranf int2}
\end{eqnarray}

we can thus employ the experimentally measured \textit{ transformed intensity }  to classify patterns. We employed a repository of $P=$ 5000 patterns containing digits with random orientation (\url{https://it.mathworks.com/help/deeplearning/ug/data-sets-for-deep-learning.html}), labeling as recognized patterns, the ones producing a transformed intensity above 10 standard deviations from the values obtained probing randomly generated binary patterns.  The value $C\_ERR$ is the number  of wrongly identified patterns experimentally.

The word experimentally here indicates that the \textit{transformed intensity} is obtained experimentally optically presenting the pattern to our \textit{disordered classifier}. The step-by-step presentation procedure is the following: \textit{i)} the probe pattern $\phi$ is printed onto a propagating laser beam employing a DMD in binary phase modulation mode (\textbf{see experimental section}), \textit{ii)} light scattered by the disordered medium is retrieved for the relevant mode/pixel set $\bm{\mathcal{M}}$, \textit{iii)} the transformed intensity measured by the selected sensors/camera pixels is obtained with  Eq. \ref{trained transformed intensity}, \textit{iv)} a pattern is defined as recognized if  the
%ref{trained transformed intensity}
\emph{trained transformed intensity}
results higher than the threshold.
The  \textit{Classification Error Probability} is then obtained as
\begin{eqnarray}
 \text{\textit{Classification Error Probability} =} C\_ERR/P.
\end{eqnarray}

\subsection{Stochastic emergent storage protocol details}
In SES we  exploit the fact that any optical coupling matrix $\bm{V^\nu}$ has a bi-dyadic thus hosting two intrinsic but random  patterns:

\begin{eqnarray}
\bm{V^\nu}=  \bm{\xi}^\nu\otimes\bm{\xi}^{\nu \dag}+\bm{\eta}^\nu\otimes\bm{\eta}^{\nu \dag}
 \end{eqnarray}

thus the optical coupling matrix at location $\nu$, $\bm{V^\nu}$, hosts the two random memory vectors $\bm{\xi^\nu}$ and $\bm{\eta^\nu}$.

To employ these disorder-embedded structures as memories we resorted to the following multi step strategy. \\

\textit{i)} We start measuring the transmission matrices from a large set of modes employing the Complete Couplings Mapping Method (CCMM, see below). We monitor $M^L=$ 65536 modes employing a region of interest for the camera of 256$\times$256 pixels in the one-mode-per-pixel configuration. The retrieved transmission matrices are saved into a computer memory and compose our starting random structures repository $\bm{\mathcal{M}}^L$.\\

\textit{ii)} We computattionally find the first eigenvector $\bm{\xi}^\nu$ for each  measured  matrix $\bm{V}^\nu$ \\

\textit{iii)}  The user, desing a target memory pattern to be stored $\bm{\phi}^*$ and a number ${M}^*$ of modes to be employed.\\

\textit{iv)}  The  target pattern  $\bm{\phi}^*$ is compared with all the eigenvectors in $\bm{\mathcal{M}}^L$ by computing the degree of similitude $\mathcal{S}$:

\begin{eqnarray}
\mathcal{S}^\nu=\hat{\bm{\phi}}^*\cdot \hat{\bm{\xi}}^\nu
\label{similitude}
\end{eqnarray}
with the symbol $\hat{\bm{i}}$ indicating  vector normalization: $\hat{i}\cdot \hat{i}=1$.

\textit{v)}  The set of modes $\bm{\mathcal{M}}^L$ is similarity-decimated to the set $\bm{\mathcal{M}}^*$, i.e. we select the  ${{M}}^*$ modes with the higher $\mathcal{S}^\nu$ values to be part of the new, reduced repository $\bm{\mathcal{M}}^*$.

Once the repository of modes is selected, we need to ``train'' the attenuation coefficients $\bm\lambda$. The attenuation values are selected between  16  values degrees of absorption in the $\in [0,1]$ range , so that they are identified with a 4 bits number.

After initializing  the lambda and computing the initial configuration optical operator
\begin{eqnarray}
\bm{J}^{\bm{\mathcal{M}}^*,\bm{\lambda}}=\sum_\nu^{{{M}}^*} \lambda^{\nu} \bm{V}^{\nu}
\label{SHS tranf int}
\end{eqnarray}
the $\bm\lambda$ are optimized with a Monte Carlo algorithm. At each optimization step a single $\lambda^{\nu}$ is modified and the change is accepted if the eigenvector similarity function

\begin{eqnarray}
\mathcal{F}^*(\bm{\lambda},\bm{\phi^*},\mathcal{\bm{M}}^*)=\hat{\bm{\xi}}_\Sigma\cdot\hat{\bm{\phi^}*}
\label{SES function}
\end{eqnarray}

decreases. Note that in Eq. \ref{SES function}, $\bm{\xi}_\Sigma$ is the first eigenvector of $\bm{J}^{\bm{\mathcal{M}}^*,\bm{\lambda}}$.

After a sufficiently large number of steps  $\mathcal{F}^*(\bm{\lambda},\bm{\phi^*},\mathcal{\bm{M}}^*)$  is minimized and form the final configuration of $\bm{\lambda}$  we obtain the final version of the optical operator:
$\bm{J}_{\bm\phi^*}^{\bm{\mathcal{M}}^*,\bm{\lambda}}$.

\subsection{Experimental Setup and CCMM}
\begin{figure}[h]
\includegraphics[width=8 cm]{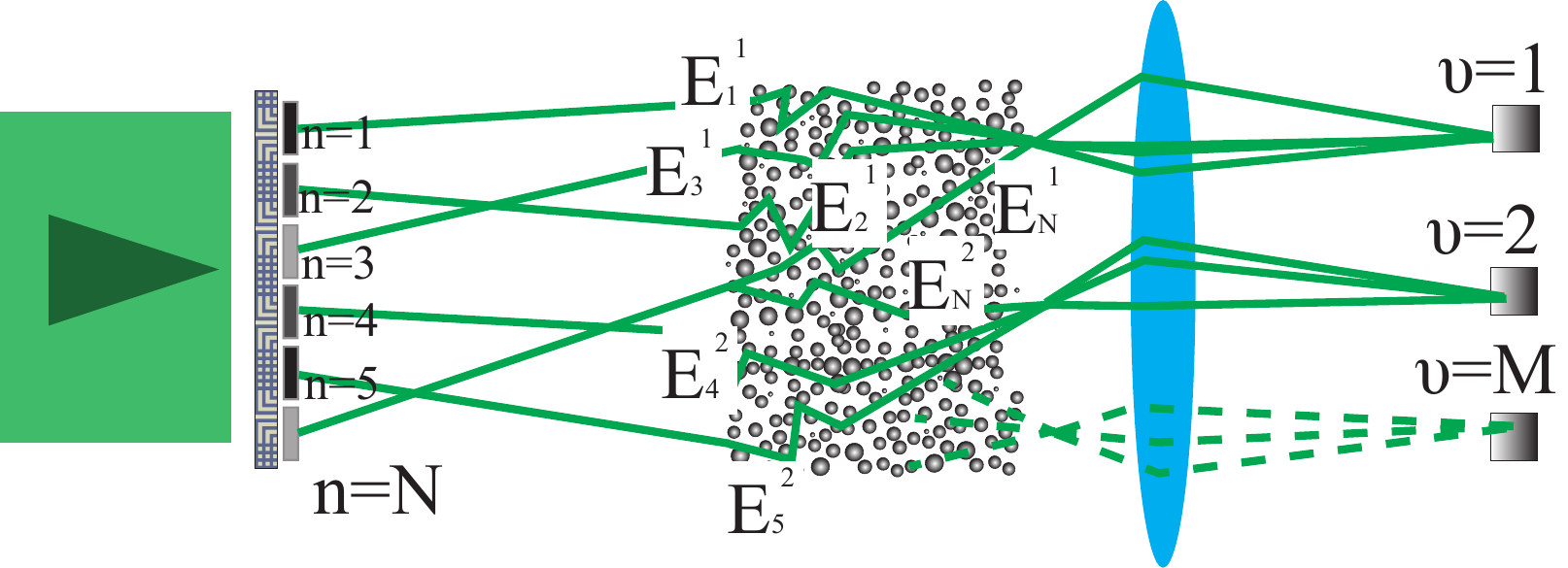}
\caption{Sketch of the experimental setup including naming for the input mode/segments ($n$ index) and detectors/output modes $\nu$ index }
\label{sketch}
\end{figure}

The same experimental setup is employed for two tasks. The first is the measurement of the optical synaptic matrices $\bm{V}^\nu$, the second is to perform classification, presenting to the disordered classifier a test pattern $\bm{\phi}$ and retrieving the \textit{transformed intensities} for each trained memory. a sketch of the experimental setup is provided in Fig. \ref{sketch}.

We employ a single mode laser (AzurLight 532, 0,5W) with beam to about 1cm. Then it is fragmented into $N$ individually modulated light rays controlled by a Digital Micromirror Device (DMD) \cite{ayoub2021high} composed by 1024$\times$768 (Vialux, V-7000, pixel Pitch 13.68 $\mu$m, 22kHz max frame rate)  flipping mirrors which can be tuned into two configurations (on or off).
Phase modulation is  obtained employing the super-pixel method (see \cite{goorden2014superpixel, leonetti2021optical} ) which require a spatial filtering to isolate the selected diffraction orders. DMD pixels are organized into 4-elements super-pixels (segments) capable to produce a 0 or $\pi$ phase pre-factors equivalent to field multiplication by $\phi_n=\in\{-1,1\}$.
The bundle of light rays is then scrambled by a diffusive, multiple scattering medium ( 60 $\mu$ layer of ZnO obtained from ZnO powder from Sigma Aldrich item 544906-50g, transport mean free path 8 $\mu$m \cite{leonetti2021spatial} ). Indeed, the DMD surface is imaged onto the Diffusive medium (0.3 $\times$ de-magnification). Then, the back layer of the disordered structure is imaged on the detection camera (11 $\times$ magnification). This magnification has been chosen to minimize the speckle size so to  work the in the one-mode-per-pixel configuration.

Superpixel method is obtained thanks to 2.66 mm aperture iris in front to the DMD.
As shown in Eq. \ref{twomirror} when two DMD mirrors are activated:
 \begin{eqnarray}
I_{n,m}^\nu = E^{\nu 2}_n+E^{\nu 2}_m+2\xi^\nu_n\xi^\nu_m+2\eta ^\nu_n\eta ^\nu_m .
\label{twomirror2}
 \end{eqnarray}
while  a single segment is activated

\begin{eqnarray}
I_n^\nu = E^{\nu 2}_n.
\label{onemirror}
\end{eqnarray}

Thus putting together Eq. \ref{twomirror2} and Eq. \ref{onemirror} one obtains

\begin{eqnarray}
V^\nu_{nm}=\xi^\nu_n\xi^\nu_m+\eta ^\nu_n\eta ^\nu_m = \frac{I_{nm}^\nu -I_{n}^\nu -I_{m}^\nu }{2}.
\label{twomirror2}
\end{eqnarray}

Thus to determine one single element of the optical synaptic matrix, one has to perform three intensity measurements. The total number of measurement to reconstruct the full synaptic matrix is $\Pi=N(N-1)/2$ (as $V^\nu_{nm}=V^\nu_{mn}$ i.e. the optical synaptic matrix is symmetric then just the above-the-diagonal elements need to be measured).
For $N$=256 this means that 32896 measurements are required, which can be obtained in maximum 5 minutes employing our DMD-Camera experimental setup (speed bottleneck from the camera sensor which works at $\sim$ 150 frames per second).
At each measurement we take a image from a Region Of Interest (ROI) of 256$\times$256 pixels, thus collecting info for $M^L=$ 65536 modes, thus generating our modes and optical synaptic matrix database. Experimental data are organized into a $32896\times256\times256$ matrix. In our case increasing the size of $N$ or $M^L$ is limited by the size of the Random Access Memory size of the computing workstation.

% Produces the bibliography via BibTeX.

\end{document}